\documentclass[pre,aps,superscriptaddress,showpacs]{revtex4-1}
\usepackage{natbib}
\usepackage{amsmath}
\usepackage{epsfig}
\usepackage{color}

\DeclareMathOperator*{\sumint}{%
\mathchoice%
{\ooalign{$\displaystyle\sum$\cr\hidewidth$\displaystyle\int$\hidewidth\cr}}
{\ooalign{\raisebox{.14\height}{\scalebox{.7}{$\textstyle\sum$}}\cr\hidewidth$\textstyle\int$\hidewidth\cr}}
{\ooalign{\raisebox{.2\height}{\scalebox{.6}{$\scriptstyle\sum $}}\cr$\scriptstyle\int$\cr}}
{\ooalign{\raisebox{.2\height}{\scalebox{.6}{$\scriptstyle\sum$}}\cr$\scriptstyle\int$\cr}}
}

\begin{document}

\title{Deriving a kinetic  uncertainty relation for piecewise deterministic processes: from classical to quantum }
\author{Fei Liu}
\email[Email address: ]{feiliu@buaa.edu.cn}
\affiliation{School of Physics, Beihang University, Beijing 100191, China}

\date{\today}

\begin{abstract}
{From the perspective of Markovian piecewise deterministic processes (PDPs), we investigate the derivation of a kinetic uncertainty relation (KUR), which was originally proposed in Markovian open quantum systems. First, stationary distributions of classical PDPs are explicitly constructed. Then, a tilting method is used to derive a rate functional of large deviations. Finally, based on an improved approximation scheme, we recover the KUR. These classical results are directly extended to the open quantum systems. We use a driven two-level quantum system to exemplify the quantum results.    }
\end{abstract}

\maketitle

\section{Introduction}
\label{section1}
{ Recently, thermodynamic uncertainty relations (TURs)~\cite{Barato2015,Pietzonka2016,Pietzonka2017,Pietzonka2018,Bertini2015,Gingrich2016,Horowitz2017,
Gingrich2017a,Proesmans2017,Barato2018,
Dechant2018,Barato2019,Hasegawa2019a, Liukq2020,VanVu2020,Koyuk2020,Manikandan2020,VanTT2020a} and kinetic uncertainty relations (KURs)~\cite{Garrahan2017,DiTerlizzi2019} of classical  nonequilibrium systems have attracted considerable interest. These generally valid relations place constraints on the precisions of time-extensive trajectory observables in terms of entropy productions~\cite{Barato2015,Gingrich2016} and dynamical activities~\cite{Garrahan2017,DiTerlizzi2019}, respectively. The studied classical systems include pure jump processes with steady-states~\cite{Barato2015,Gingrich2016,Garrahan2017} or for cases of finite time~\cite{Liukq2020}, periodically driven jump processes with quasi-steady states ~\cite{Barato2018,Barato2019}, and continuous diffusions~\cite{Hasegawa2019a,Dechant2018}.

With establishments of the classical results, some efforts have been devoted to the quantum regime~\cite{Agarwalla2018,Ptaszy2018,Hasegawa2019,Liu22019,Guarnieri2019,Carollo2019,Hasegawa2020,
Carollo2021,Menczel2021}. Quantum physics manifests some novel features for the classical relations. For instance, Agarwalla and Segal~\cite{Agarwalla2018} demonstrated that quantum coherence systems that do not follow a population Markovian master equation can violate the classical TURs. For Markovian open quantum systems, by using quantum estimation theory, Hasegawa obtained a quantum TUR that holds for arbitrary continuous measurements satisfying a scaling condition. Interestingly, the bound in the relation is sensitive to measurement schemes~\cite{Hasegawa2020}. For the same type of open quantum systems, a KUR~\footnote{In their original paper, Carollo et al. called their uncertainty relation as a generalized TUR. Because their bound depends on quantum-corrected dynamic activity instead of thermodynamic entropy production (see Eq.~(\ref{generalizedTURPDPs}) below) we think that a more appropriate term is KUR~\cite{DiTerlizzi2019}. Because the KUR also holds for many classical systems, which will be demonstrated in this paper, we do not overemphasize its quantum character. } was proposed by Carollo et al.~\cite{Carollo2019}. They found that this quantum relation can yield higher precision than that of the classical counterpart~\cite{Garrahan2017}. The physical phenomena underlying this include the famous antibunching effects revealed in quantum optics~\cite{Mandel1995}.

{In this paper, we reinvestigate the derivation of the KUR found by Carollo et al. in Markovian open quantum systems~\cite{Carollo2019}. There are two motivations to carry out this effort. On one hand, we want to show that the KUR actually holds for a general type of stochastic processes, the Markovian piecewise deterministic processes (PDPs)~\cite{Davis1984,Davis1993}, either in classical or in quantum regimes. Hence, the improved precision of the KUR found by Carollo et al. is not unique to quantum physics. On the other hand, we attempt to make the previous derivation concise and clear. Compared with the establishment of the classical KUR~\cite{Garrahan2017}, the quantum extension is far more complicated, since the Liouville-master equation in Hilbert space~\cite{Breuer2002} was used. We think that this complexity could be avoidable if trajectory method for the PDPs is exploited.  }

To achieve the aim, we focus on the derivation of a KUR for the PDPs of classical nonequilibrium systems, while the case of open quantum systems is treated as natural extension of the classical relation. Technically, we use probability formulas of trajectories of the PDPs. We show that their applications simplify the construction of the stationary distributions of the PDPs and the following derivation of the KUR. The underlying foundation of realizing the above two points is that rates of jumps, rather than classical stochastic variables or quantum wave functions that describe states of classical or quantum systems, directly determine fluctuation structures of the PDPs. This observation is the most important insight achieved in this paper. Before we start the main text, we emphasize that several crucial formulas re-derived here must be credited to Carollo et al.~\cite{Carollo2019}, and most derivations performed here are heuristic; Mathematical rigor is not our concern.

The remainder of this paper is organized as follows. In Sec.~(\ref{section2}), after defining the PDP of a classical system, we derive the Liouville-master equation and its stationary distribution. In Sec.~(\ref{section3}), we use a tilting method to obtain a key rate functional of large deviations (LDs). Based on the result, in Sec.~(\ref{section4}) we recover the KUR of Carollo et al.~\cite{Carollo2019} for the classical PDP. In Sec.~(\ref{section5}), we briefly extend these classical results to the open quantum system situation. A simple two-level quantum system is used to exemplify our quantum extensions. Section~(\ref{section6}) concludes the paper.

\section{PDPs of classical systems}
\label{section2}
We begin with the PDP of a classical system~\cite{Davis1984,Davis1993}. Real examples include membrane voltage fluctuations in a neuron~\cite{Fox1994,Keener2011}, some biochemical processes like molecular motors and gene regulation~\cite{Faggionato2010}; also see the literatures in Ref.~\cite{Bressloff2014}. Let the terminal time of the process be $T$. A concrete realization of the stochastic process is named as trajectory. For simplicity of notation, we use a one-dimensional continuous variable $x$ to denote the state of the stochastic system. The PDP is composed of deterministic pieces and stochastic jumps. Between jumps, the dynamics of the system is governed by a deterministic drift, e.g., $g(x)$~\footnote{In this paper we exclude the special case of zero drift.}. The corresponding equation of motion is
\begin{eqnarray}
\label{determinsiticdifferentialequation}
\frac{\rm d}{{\rm d}\tau}x(\tau)=g(x(\tau)),
\end{eqnarray}
where time zero of $\tau$ is set immediately after the last jump. Because the system has a chance to jump from the current state to others, these deterministic pieces are randomly interrupted at discrete time points. We denote the rates of these jumps as $w_\alpha(x|y)$, where $\alpha=1,\cdots, M$ indicate the types of the jumps, $M$ is set to a fixed number, and $x$ and $y$ are the states of the system before and after the jumps, respectively. If a jump of type $\alpha$ happens, we simply call it $\alpha$-jump. In this paper, we additionally restrict the states after jumps to be discrete and to belong to a fixed set. Because they are accessed by certain types of jumps, we call these states target states and denote them $x_\alpha$. Accordingly, the initial condition of Eq.~(\ref{determinsiticdifferentialequation}) is one of the target states. We express the flow or the solution of the deterministic equation as $\phi_{\alpha}(\tau)$. This restriction not only simplifies our discussions, but also is consistent with the case of open quantum systems, where quantum systems collapse to one of finite wave functions after each jumps~\cite{Breuer2002}; also see Eq.~(\ref{targetwavefunction}) below. Note that $x_\alpha=x_\beta$ is allowed for different $\alpha$- and $\beta$-jumps.

With the rates mentioned above, we can obtain the survival time distribution function that the system successively evolves without jumps after the last jump~\cite{Breuer2002}. Let the latter be $\alpha$ and the target state be $x_\alpha$. Then, the time distribution function of successive evolution $S_\alpha(\tau)$ satisfies
\begin{eqnarray}
\label{differentialeqSa}
\frac{{\rm d}}{{\rm d}\tau}S_\alpha(\tau)&=&-\sum_{\beta=1}^M S_\alpha(\tau)w_\beta(\phi_{\alpha}(\tau)|x_\beta).
\end{eqnarray}
Obviously, the term in the sum of Eq.~(\ref{differentialeqSa}) is the conditional probability density of the next $\beta$-jump happening at time $\tau$ given the last $\alpha$-jump. We specially denote it as
\begin{eqnarray}
\label{conditionalproba|b}
p_{\alpha|\beta}(\tau)=S_\alpha(\tau)w_\beta(\phi_{\alpha}(\tau)|x_\beta).
\end{eqnarray}

Defining the total rate
\begin{eqnarray}
\Gamma(x)=\sum^M_{\beta=1} w_\beta(x|x_\beta)\equiv\int {\rm d}y W(x|y),
\end{eqnarray}
where
\begin{eqnarray}
\label{additionaldefinitionrate}
{W}(x|y)=\sum^M_{\beta=1} w_\beta(x|x_\beta)\delta(y-x_\beta)
\end{eqnarray}
and the integration $\int {\rm d} y$ is over the whole space of the stochastic variable $x$,
we obtain
\begin{eqnarray}
\label{survivalprob}
S_{\alpha}(\tau)={\rm e}^{-\int_0^\tau {\rm d}s\Gamma(\phi_{\alpha}(s))}.
\end{eqnarray}
The definition of $W(x|y)$ is not essential, but it makes the Liouville-master equation more compact in form; see Eq.~(\ref{Liouvillemasterclassicalsystem}) below.

\subsection{Liouville-master equation}
The conventional approach of studying stochastic processes is to establish the integral of differential Chapman-Kolmogorov equations~\cite{Risken1984}. For the PDPs, the corresponding equation for the probability density function $p(x,t)$ is known~\cite{Davis1984,Breuer2002}. In the following, we derive the same result by an intuitive age-structure formalism used in semi-Markov processes~\cite{Ross1995,Wang2007}. This enables us to construct the stationary distribution of the PDPs.

Let $p(x,t;\alpha,\tau)$ be the joint probability density when the state of the system at time $t$ is $x$, the time of successive evolution is $\tau$, and the type of the last jump is $\alpha$. This function has several simple properties. Firstly, it satisfies a normalization condition:
\begin{eqnarray}
\label{normalizationcondition}
\int {\rm d}x \left[\sum_{\beta=1}^M  \int_0^\infty {\rm d}\tau p(x,t;\beta,\tau)\right]=1.
\end{eqnarray}
Note that the complete term in the squared brackets of Eq.~(\ref{normalizationcondition}) is the probability density function $p(x,t)$ of the system in state $x$ at time $t$. Secondly, the chance of the system evolving without jumps continuously decreases with time, as the time $\tau$ approaches infinity, the function $p(x,t;\alpha,\tau)$ decays to zero; also see Eq.~(\ref{survivalprob}). Finally, considering that $p(x,t;\alpha,\tau=0)$ is also equal to the probability density that other continuous states jump to $x_\alpha$ through $\alpha$-jump at time $t$, while these states may start with jumps of different types with various times of continuous evolution, we can express it as
\begin{eqnarray}
\label{initialconditionprocess}
p(x,t;\alpha,0)&=&\sum_{\beta=1}^M\int {\rm d}y\int_0^\infty {\rm d}\tau p(y,t;\beta,\tau) w_\alpha(y|x_\alpha) \delta (x-x_\alpha)\nonumber \\
& \equiv& c_\alpha(t)\delta(x-x_\alpha).
\end{eqnarray}
The defined functions $c_\alpha(t)$, $\alpha=1,\cdots, M$, in the second line of Eq.~(\ref{initialconditionprocess}) represent the rates of $\alpha$-jumps happening at time $t$. We will give their concrete expressions in the steady state of the PDP. Note $\delta(x-
x_\alpha)$ in Eq.~(\ref{initialconditionprocess}) is essential since the state of the system $x$ must equal $x_\alpha$ immediately after the $\alpha$-jump.

Let $h$ be a small time interval. We have
\begin{eqnarray}
\label{probrelationbetweentwotimes}
p(x,t+h;\alpha,\tau+h)=\int {\rm d}y p(y,t;\alpha,\tau)\left[1-\Gamma(y)h\right]\delta(x-y- g(y)h)+{\it o}(h).
\end{eqnarray}
The reason for this is that if the system starts with the target state $x_\alpha$ after the $\alpha$-jump and successively evolves $\tau+h$ at time $t+h$, the system must evolve $\tau$ at time $t$, and no jumps occur during $h$. Note that the Dirac function $\delta()$ therein indicates the deterministic Eq.~(\ref{determinsiticdifferentialequation}). Expanding both sides of Eq.~(\ref{probrelationbetweentwotimes}) in terms of $h$ and letting $h\rightarrow 0$, we obtain
\begin{eqnarray}
\label{subLiouvillemasterequation}
\partial_t p(x,t;\alpha,\tau)+\partial_\tau p(x,t;\alpha,\tau)=-\partial_x (g(x)p(x,t;\alpha,\tau) )- \Gamma(x)p(x,t;\alpha,\tau).
\end{eqnarray}
Integrating and summing Eq.~(\ref{subLiouvillemasterequation}) with respect to $\tau$ and $\alpha$, respectively, and using the properties of $p(x,t;\alpha,\tau)$ at $\tau=0$ and $\infty$, respectively, we obtain the classical Liouville-master equation for the probability density function of the PDP~\cite{Davis1984,Breuer2002}:
\begin{eqnarray}
\label{Liouvillemasterclassicalsystem}
\partial_t p(x,t)&=&-\partial_x (g(x)p(x,t))+\int {\rm d}y [p(y,t)W(y|x)-p(x,t)W(x|y)].
\end{eqnarray}
Now, the reason for defining $W(x|y)$ [Eq.~(\ref{additionaldefinitionrate})] is clear.

\subsection{Stationary distribution of PDPs}
Let us assume that the classic system has a unique steady state over long time durations~\footnote{Davis has discussed conditions for the existence of steady states of PDPs; see p.127 in Ref.~\cite{Davis1993}. }. Under this situation, $p(x,t=\infty;\alpha,\tau)\equiv p_\infty(x;\alpha,\tau)$ is independent of time $t$. Using the flow of Eq.~(\ref{determinsiticdifferentialequation}), we formally solve Eq.~(\ref{subLiouvillemasterequation}):
\begin{eqnarray}
\label{distributionsteadystate}
p_\infty(x;\alpha,\tau)=c_\alpha S_\alpha(\tau)\delta(x-\phi_{\alpha}(\tau)).
\end{eqnarray}
Note that here $c_\alpha$ is also no longer time-dependent. Substituting the solution into Eq.~(\ref{initialconditionprocess}) and integrating both sides with respect to $x$, we have
\begin{eqnarray}
\label{eigenvector}
c_\alpha&=&\sum_{\beta=1}^M c_\beta \left[ \int_0^\infty {\rm d}\tau p_{\beta|\alpha}(\tau)\right] \nonumber \\
&\equiv  &\sum_{\beta=1}^M c_\beta P_{\beta|\alpha }.
\end{eqnarray}
According to the probability meaning of $p_{\beta|\alpha}(\tau)$, we see that $P_{\beta|\alpha}$ is the conditional probability that the type of the next jump is $\alpha$ given the last $\beta$-jump, and obviously $\sum_{\alpha=1}^M P_{\beta |\alpha }=1$. Therefore, $P_{\beta|\alpha}$ is the transition probability of the embedded Markov chain of the PDP, while $c_\alpha$ is proportional to the stationary distribution of the Markov chain~\cite{Ross1995}. Let $\pi_\alpha$, $\alpha=1,\cdots,M$ be the distribution. According to Eq.~(\ref{normalizationcondition}), we have
\begin{eqnarray}
\label{ratealpha}
c_\alpha=\frac{\pi_\alpha}{\sum_{\beta=1}^M \pi_\beta\tau_\beta},
\end{eqnarray}
where
\begin{eqnarray}
\label{meansurvivaltime}
\tau_\beta=\int_0^\infty {\rm d}\tau S_\beta(\tau)=\sum_{\gamma=1}^M\int_0^\infty {\rm d}\tau p_{\beta|\gamma}(\tau)  \tau
\end{eqnarray}
is the average time of continuous evolution of the system starting with the target state $x_\beta$ of the last $\beta$-jump. Eq.~(\ref{ratealpha}) clearly shows that $c_\alpha$ is indeed the rate of $\alpha$-jump in the steady state. If we further use $p_\infty(x)$ to denote the probability density function $p(x,t)$ under the long time limit, the relationship between $p(x,t)$ and $p(x,t;\beta,\tau)$ reminds us
\begin{eqnarray}
\label{stationarydensity}
p_\infty(x)=\sum_{\beta=1}^M c_\beta \int_0^\infty {\rm d}\tau S_\beta(\tau) \delta(x-\phi_\beta(\tau) ).
\end{eqnarray}
The probability meaning of Eq.~(\ref{stationarydensity}) is intuitive.

\section{A rate functional of large deviations}
\label{section3}
In this section, we use a tilting method~\cite{Maes2008a,Maes2008,Barato2015a} to derive a rate functional  of LDs of the PDPs. Quantum extension leads to the KUR of open quantum systems~\cite{Carollo2019}. To this end, we write the probability functional ${\cal P}[X_T]$ of a trajectory $X_T$ of the PDP. We assume that in the trajectory  a sequence of $\alpha_i$-jumps happen at times $t_i$, where $i=0,1,\cdots, N$, and $N$ is the total number of jumps. Furthermore, we denote the states of the system just before and immediately after the time $t_i$ as $x(t_i^-)$ and $x(t_i^+)$, respectively. According to our previous notations, $x(t_i^+)$ is equal to one of the target states: that is, $x(t_i^+)=x_{\alpha_i}$, while $x(t_i^-)$ is equal to $\phi_{\alpha_{i-1}}(\tau_i)$, where $\tau_i=t_{i}-t_{i-1}$. A schematic diagram of such a trajectory and concrete explanations of these notations are shown in Fig.~(\ref{fig1}). Then, the probability functional of the trajectory is
\begin{eqnarray}
\label{probdistributiontraj}
{\cal P}[X_T]
=p_{\alpha_0|\alpha_1}(\tau_1)p_{\alpha_1|\alpha_2}(\tau_2)\cdots p_{\alpha_{N-1}|\alpha_N}(\tau_N)S_{\alpha_N}(T-t_N).
\end{eqnarray}
Using the Dirac delta function and Kronecker delta symbol, we rewrite Eq.~(\ref{probdistributiontraj}) as
\begin{eqnarray}
\label{probdistributiontraj2}
{\cal P}[X_T]
\asymp {\rm e}^{ T \sum_{\beta,\gamma=1}^M\int_0^\infty {\rm d}\tau \ln p_{\beta|\gamma}(\tau)F_{\beta\gamma;\tau}[X_T]},
\end{eqnarray}
where
\begin{eqnarray}
\label{Ffunctionaldefinition}
F_{\beta\gamma;\tau}[X_T]\equiv \frac{1}{T}\sum_{i=1}^N \delta_{\beta,\alpha_{i-1}}\delta_{\gamma,\alpha_{i}}\delta(\tau_i-\tau).
\end{eqnarray}
Because this paper pertains to the situation of a long time limit, we use ``$\asymp$" to indicate that the initial probability of the system and the last term $S_{\alpha_N}$ in Eq.~(\ref{probdistributiontraj}) are neglected. Eq.~(\ref{Ffunctionaldefinition}) represents the average number of the ordered $\beta$- and $\gamma$-jumps next to each other in unit time along the trajectory $X_T$ and their interval equal to $\tau$. If the classic system is ergodic, under the long time limit the law of large numbers almost surely ensures that $F_{\beta\gamma;\tau}$ converges to $c_\beta p_{\beta|\gamma}(\tau)$.
\begin{figure}
\includegraphics[width=1.\columnwidth]{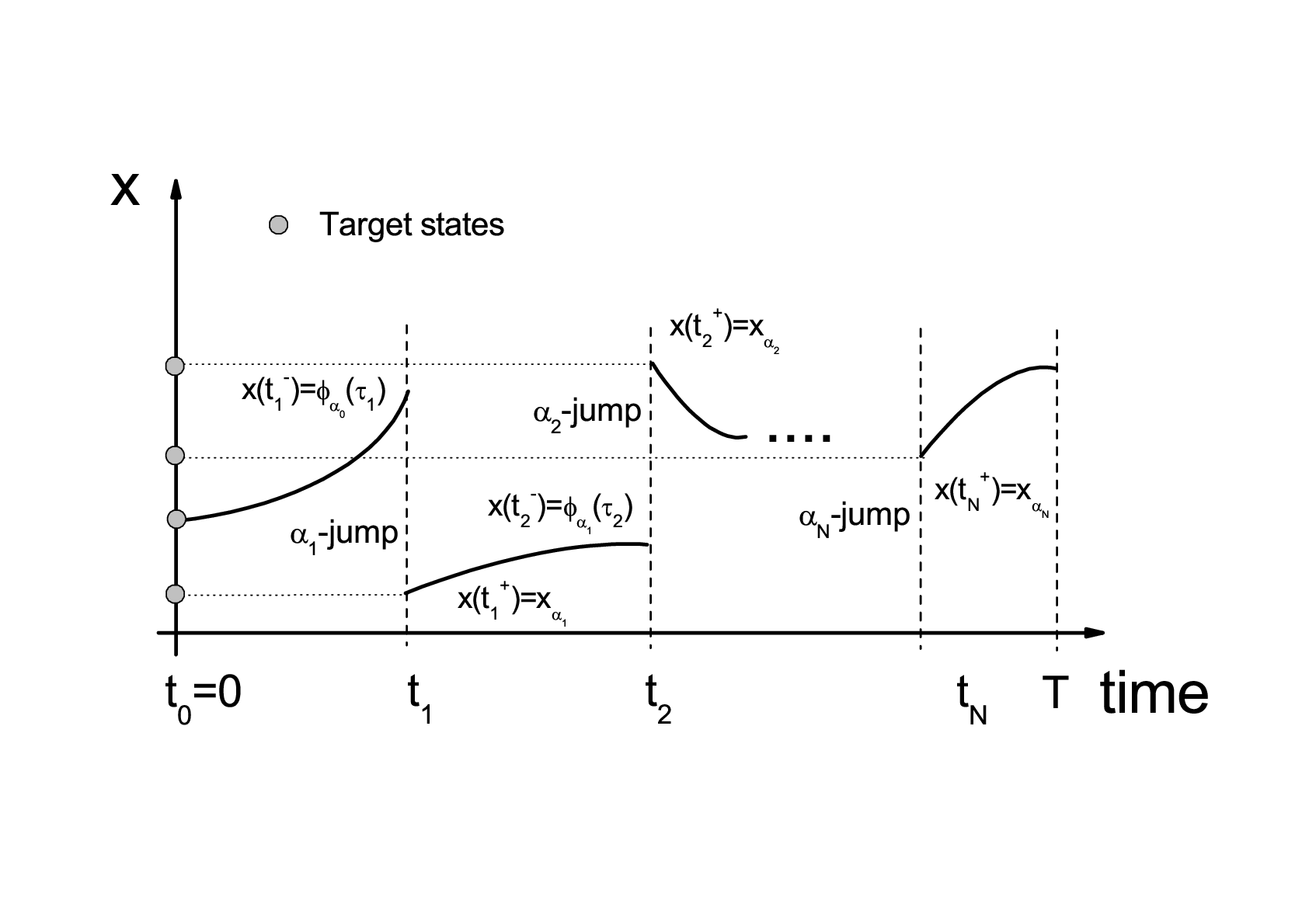}
\caption{A schematic diagram of a trajectory of the classical PDP.    }
\label{fig1}
\end{figure}

Let $p[F]$ be the probability distribution functional for the functions $F_{\alpha\beta}(\tau)$, $\alpha,\beta=1,\cdots,M$. Based on Eq.~(\ref{probdistributiontraj2}),  we have
\begin{eqnarray}
\label{probdistributionpF}
p[F]
&\asymp& \sum_{X_T} {\cal P}[X_T]\prod_{\beta,\gamma=1}^M\prod_{j=1}^K\delta(
F_{\beta\gamma;s_j}[X_T] -F_{\beta\gamma}(s_j))
\nonumber \\
&\asymp &\sum_{X_T} \widetilde{{\cal P}}[X_T]
\frac{{\cal P}[X_T]}
{\widetilde{{\cal P}}[X_T]}\prod_{\beta,\gamma=1}^M\prod_{j=1}^K\delta(
F_{\beta\gamma;s_j}[X_T] -F_{\beta\gamma}(s_j))\nonumber\\
&\asymp& \exp\left[-T \sum_{\beta,\gamma=1}^M\int_0^\infty {\rm d}\tau F_{\beta\gamma}(\tau)\ln \frac{\widetilde {p}_{\beta|\gamma}(\tau)}{p_{\beta|\gamma}(\tau)}\right] \sum_{X_T} \widetilde{{\cal P}}[X_T]\prod_{\beta,\gamma=1}^M\prod_{j=1}^K\delta(
F_{\beta\gamma;s_j}[X_T] -F_{\beta\gamma}(s_j)),\nonumber\\
\end{eqnarray}
where $s_j=j\epsilon$, $j=1,\cdots,K$ and $K=T/\epsilon$, and the sum is over all trajectories of the PDP. Note that the right-hand side of Eq.~(\ref{probdistributionpF}) must be understood in the limit $K\rightarrow\infty$, since $\tau$ is a continuous variable. In the second equation, we introduce a fictitious classical system and its dynamical description is still PDP: in this system, the probability density functional of the trajectory $X_T$ is $\widetilde{{\cal P}}[X_T]$, and $\widetilde{ p}_{\beta|\gamma}(\tau)$ is the conditional probability density, like $p_{\beta|\gamma}(\tau)$ of the real system. Apparently, the last sum of the third equation is exactly the probability density functional $\widetilde{p}[F]$ of the fictitious system. Importantly, if $F_{\beta\gamma}(\tau)$ is typical in the steady state of this system, that is, $\widetilde{p}[F]\simeq 1$, Eq.~(\ref{probdistributionpF}) leads to a rate functional of the original system:
\begin{eqnarray}
\label{ratefunctionalFunctions}
I[F]=\sum_{\beta,\gamma=1}^M\int_0^\infty {\rm d}\tau F_{\beta\gamma}(\tau)\ln \frac{\widetilde {p}_{\beta|\gamma}(\tau)}{p_{\beta|\gamma}(\tau)}.
\end{eqnarray}
where
\begin{eqnarray}
\label{condition}
F_{\beta\gamma}(\tau)=\widetilde{c}_\beta\widetilde{p}_{\beta|\gamma}(\tau),
\end{eqnarray}
and
\begin{eqnarray}
{\widetilde c}_\beta=\sum_{\gamma=1}^M\int_0^\infty {\rm d}\tau F_{\beta\gamma}(\tau)
\end{eqnarray}
is simply the rate of $\beta$-jump in the steady state of the fictitious system; see Eq.~(\ref{ratealpha}). Eq.~(\ref{ratefunctionalFunctions}) agrees with that obtained by Carollo et al. in Markovian open quantum systems~\cite{Carollo2019}. Note that so far our discussions are restricted to the classical regime. We explain below that this is not accidental.

\section{A kinetic uncertainty relation}
\label{section4}
Following Carollo et al.~\cite{Carollo2019}, the time-extensive trajectory observable we are concerned with is a linear combination of the empirical rates ${k}_\alpha$
which is the number of $\alpha$-jump happening in unit time along a trajectory $X_T$:
\begin{eqnarray}
\label{stochasticcurrent}
{J}[X_T]=\sum_{\alpha=1}^M \omega_\alpha {k}_\alpha[X_T].
\end{eqnarray}
Here, $\omega_\alpha$, $\alpha=1,\cdots,M$, represent certain $\alpha$-dependent constants. That is, they are specified by the types of the jumps.
We emphasize that these $k_\alpha$ are functionals of trajectory. Even if the system has been in the steady state, they usually do not equal $c_\alpha$ because of fluctuations in finite time interval $T$. Of course, observing these large fluctuations is rare, and their values are mainly surrounding $c_\alpha$. Accordingly, the observed $J$ is fluctuating around the typical
\begin{eqnarray}
{J}_{s}=\sum_{\alpha=1}^M \omega_\alpha c_\alpha.
\end{eqnarray}
To quantify their deviation, we introduce a dimensionless parameter $\lambda= 1+\eta$: ${J}=\lambda {J}_{s}$. Obviously, $\eta$ is equal to $({J}-{J}_{s})/{J}_{s}$. Here, we do not consider the cases of $J_s=0$.

Assume that the LD principle is valid for the observable~(\ref{stochasticcurrent}): that is, the probability density function for $J$ is $ p({J})\asymp \exp[-TI(J)]$~\cite{Touchette2008}. According to the contraction principle~\cite{Touchette2008}, the rate function $I(J)$ is equal to the minimum of $I_M(k_1,\cdots,k_M)$:
\begin{eqnarray}
\label{ratefunctionofcurrent}
I({J})=\min_{J} I_M(k_1,\cdots,k_M)\le I_M(k_1,\cdots,k_M),
\end{eqnarray}
where $I_M$ is the rate function for the stochastic rates $k_\alpha$, and the minimization is achieved by selecting all possible rates that satisfy the constraint~(\ref{stochasticcurrent}). Note that on the right-hand side of the inequality in Eq.~(\ref{ratefunctionofcurrent}), $k_\alpha$ represents such a column of arbitrary rates. A simple case is $k_\alpha=\lambda c_\alpha$. Note that $\eta$ is now also equal to $(k_\alpha -c_\alpha)/{c_\alpha}$. We also do not consider the extreme cases of $c_\alpha=0$. Crucially, the rate function $I_M$ can be further obtained by applying the contraction principle again to the rate functional~(\ref{ratefunctionalFunctions}):
\begin{eqnarray}
\label{inequality2}
I_M(k_1,\cdots, k_M)=\min_{ k_1,\cdots,k_M} I[F]\le I[F].
\end{eqnarray}
Analogously, the minimization of Eq.~(\ref{inequality2}) is achieved by selecting all possible $F_{\beta\gamma}(\tau)=\widetilde c_\beta\widetilde p_{\beta|\gamma}(\tau)$ that satisfy the constraint $\widetilde c_\beta= k_\beta$, where $\widetilde c_\beta$ is defined by Eq.~(\ref{condition}). The notation $F$ on the right-hand side of the inequality in Eq.~(\ref{inequality2}) represents such functions.

We now focus on constructing concrete $F_{\beta\gamma}(\tau)$. Because of the constraint, arbitrariness of these functions is in fact inherited by ${\widetilde p}_{\beta|\gamma}(\tau)$.  Analogous to previous schemes in the classical regime~\cite{Gingrich2014,Gingrich2016,Barato2018,Barato2019}, if we are concerned with a local quadratic upper bound of $I(J)$ around ${J}_s$, that is, $\eta$ is regarded as a small quantity, we can present an ansatz that ${\widetilde p}_{\beta|\gamma}(\tau)$, the conditional probability density function of an auxiliary system, is constructed as a perturbation of $p_{\beta|\gamma}(\tau)$ of the original system:
\begin{eqnarray}
\label{ansatz}
{\widetilde p}_{\beta|\gamma}(\tau)= p_{\beta|\gamma}(\tau)\left[1+\eta f_{\beta\gamma}(\tau)+\eta^2 g_{\beta\gamma}(\tau)\right]+{\it o(\eta^2)}.
\end{eqnarray}
$f_{\beta\gamma}(\tau)$ and $g_{\beta\gamma}(\tau)$ are the two functions to be determined.

The constraint ${\widetilde c}_\beta=\lambda c_\beta$ implies two consequences. First, according to Eq.~(\ref{eigenvector}), the transition probability ${\widetilde P}_{\beta|\gamma}$ of the Markov chain of the auxiliary system must be equal to $P_{\beta|\gamma}$ of the original system. Hence, we have
\begin{eqnarray}
\label{firstresultofconstrain}
\int_0^\infty {\rm d}\tau [\eta f_{\beta\gamma}(\tau) + \eta^2 g_{\beta\gamma}(\tau) ] p_{\beta|\gamma}(\tau)=0.
\end{eqnarray}
On the other hand, when we substitute the ansatz~(\ref{ansatz}) into Eq.~(\ref{ratefunctionalFunctions}), keep the terms until the second order of $\eta$, and use Eq.~(\ref{firstresultofconstrain}), we find that the right-hand side of the inequality in Eq.~(\ref{inequality2}) equals
\begin{eqnarray}
\label{boundedratefunctional}
\frac{\eta^2}{2}\sum_{\beta,\gamma=1}^Mc_\beta\int_0^\infty  {\rm d}\tau p_{\beta|\gamma}(\tau)\left[2f_{\beta\gamma}(\tau)+f_{\beta\gamma}(\tau)^2\right].
\end{eqnarray}
The term of the first order of $\eta$ exactly vanishes. The second consequence comes from Eq.~(\ref{ratealpha}): $\widetilde\tau_\beta=\tau_\beta/\lambda$, where $\widetilde\tau_\beta$ is the average survival time of the auxiliary system. According to Eq.~(\ref{meansurvivaltime}), we expand the definition of the average survival time in terms of $\eta$ and obtain an equation with respect to the first order of $\eta$:
\begin{eqnarray}
\label{restriction2}
\int_0^\infty {\rm d}\tau p_{\beta|\gamma}(\tau)\tau=-\int_0^\infty {\rm d}\tau p_{\beta|\gamma}(\tau)f_{\beta\gamma}(\tau)\tau.
\end{eqnarray}
A simple but non-trivial solution is
\begin{eqnarray}
f_{\beta\gamma}(\tau)=
\frac{\tau_{\beta|\gamma}^2}
{\sigma_{\beta|\gamma}^2}\left(1-\frac{\tau}{\tau_{\beta|\gamma}}\right),
\end{eqnarray}
where
\begin{eqnarray}
\tau_{\beta|\gamma}&=&P_{\beta|\gamma}^{-1}\int_0^\infty {\rm d}\tau p_{\beta|\gamma}(\tau)\tau \\
\sigma_{\beta|\gamma}^2&=&P_{\beta|\gamma}^{-1}\int_0^\infty {\rm d}\tau p_{\beta|\gamma}(\tau)\tau^2- (\tau_{\beta|\gamma})^2
\end{eqnarray}
are the average and variance of the time of continuous evolution when the type of the next jump is $\gamma$ given the last $\beta$-jump, respectively. Note that $P_{\beta|\gamma}^{-1}$ therein fills the role of normalization with respect to time $\tau$. Carollo et al.~\cite{Carollo2019} defined these two quantities.

Substituting the concrete formula of $f_{\beta\gamma}(\tau)$ into Eq.~(\ref{boundedratefunctional}) and following the inequalities in Eq.~(\ref{ratefunctionofcurrent}) and~(\ref{inequality2}), we obtain an upper bound of the rate function of the stochastic observable:
\begin{eqnarray}
\label{upperboundofrate}
I({J})\le \frac{({J}-{ J}_{s})^2}{2{J}_{s}^2}\sum_{\beta,\gamma=1}^M c_\beta P_{\beta|\gamma} \frac{\tau_{\beta|\gamma}^2} {\sigma_{\beta|\gamma}^2}.
\end{eqnarray}
The last step is to consider that the variance of the stochastic observable within the finite duration $T$ is equal to $\sigma_{J}^2=1/TI''({J}_{s})$, where two primes denote the second derivative with respect to $J$, and the upper bound implies a KUR:
\begin{eqnarray}
\label{generalizedTURPDPs}
\sigma_{J}^2 \ge { J}_{s}^2 \left(T\sum_{\beta,\gamma=1}^M c_\beta P_{\beta|\gamma} \frac{\tau_{\beta|\gamma}^2} {\sigma_{\beta|\gamma}^2}\right)^{-1}.
\end{eqnarray}

{Let us make several comments. First of all, Eq.~(\ref{generalizedTURPDPs}) is a KUR instead of TUR, since the lower bound on the right-hand side involves the rates of jumps of the system and is not directly relevant to entropy production. Second, the KUR is proven for the classical PDP. Nevertheless, it is exactly the same as that in Markovian open quantum systems~\cite{Carollo2019}. Therefore, as we stated at the beginning, the KUR indeed holds for the classical systems and open quantum systems whose dynamics are described by PDPs}. Finally, Carollo et al. made an alternative approximation~\footnote{In their paper, the indices are $i$ and $j$ instead of $\alpha$ and $\beta$. The details can be found in S2.C of the supplemental material of the paper.  }: $
{\widetilde p}_{\beta|\gamma}(\tau)=v_{\beta\gamma}\exp(-\tau u_{\beta\gamma})p_{\beta|\gamma}(\tau)$,
where the parameters $u_{\beta\gamma}$ are regarded as small quantities and $v_{\beta\gamma}$ are to be determined. By contrast, since one single small parameter $\eta$ is included, the approximation~(\ref{ansatz}) leads to a simpler argument toward the KUR.

\section{Extensions to open quantum systems}
\label{section5}
When we recall the probability formulas and statistics of the KUR in previous sections, we notice that they are directly determined by the rates $w_\alpha(x|x_\alpha)$ of the jumps. Therefore, if retaining these rates while regarding $x$ as the wave function that describes the state of some quantum system, and updating the classical Eq.~(\ref{determinsiticdifferentialequation}) into some Schr$\ddot{o}$dinger-like equations, we will obtain a theory regarding the PDPs and KUR of the quantum systems. Because these PDPs exist about the wave functions in Hilbert space, their trajectories are specially coined {\it quantum} trajectories~\cite{Carmichael1993}.  Physically, the PDPs of wave functions or density matrixes arise in quantum systems that are continuously measured or monitored~\cite{Breuer2002,Wiseman2010}. In the remainder of this paper, after sketching established knowledge about the PDPs of general open quantum systems~\cite{Breuer2002}, we briefly examine the quantum extensions of the previous classical results. Finally, we exemplify them by a simple two-level quantum system.

\subsection{Unraveling the quantum master equation}
Let $\rho(t)$ be the reduced density matrix of an open quantum system. The ensemble dynamics of the system is described by the Markovian quantum master equation~\cite{Davies1974,Lindblad1976,Gorini1976,Breuer2002}
\begin{eqnarray}
\label{quantummasterequation}
\partial_t \rho(t)=-{\rm i}[H,\rho(t)]+\sum_{\alpha=1}^M r_\alpha[ A_\alpha\rho(t)A^\dag_\alpha -\frac{1}{2}\left\{A^\dag_\alpha A_\alpha,\rho(t)\right\}],
\end{eqnarray}
where the Planck constant $\hbar$ is set to 1, $H$ denotes the Hamiltonian of the quantum system, $A_\alpha$ is the jump operator or Lindblad operator and is assumed to be eigenoperator of a certain Hamiltonian of the system, and nonnegative $r_\alpha$, $\alpha=1,\cdots,M$ represent the correlation functions of the environments surrounding the system. The reduced density matrix of this general equation is demonstrated to equal to an average of the density matrixes of the individual quantum systems~\cite{Breuer1997a}:
\begin{eqnarray}
\label{densitymatrixzz}
\rho(t)=\int {\rm D}\psi {\rm D}\psi^* P[\psi,t] |\psi\rangle\langle \psi|.
\end{eqnarray}
Here, $D\psi D\psi^*$ is the Hilbert space volume element, and $P[\psi,t]$ is the probability distribution functional of the stochastic wave function $\psi$ at time $t$. The latter satisfies the quantum Liouville-master equation:
\begin{eqnarray}
\label{Liouvillemasterequationquantum}
\partial_t P[\psi,t]&=&{\rm i}\int {\rm d}z\left\{ \frac{\delta}{\delta \psi(z)}G[\psi](z)-\frac{\delta}{\delta \psi^*(z)}G[\psi]^*(z) \right\} P[\psi,t]\nonumber +\nonumber \\
&&\int {\rm D}\phi {\rm D}\phi^*\left\{ P[\phi,t]W[\phi|\psi] -P[\psi,t]W[\psi|\phi] \right\},
\end{eqnarray}
where $\delta/\delta \psi(z)$ and $\delta/\delta^* \psi(z)$ are functional derivatives, and $z$ denotes the positional representation. The operator $G$ in the first integral of Eq.~(\ref{Liouvillemasterequationquantum}) is
\begin{eqnarray}
\label{Goperator}
G[\psi]=(\hat H  + \frac{{\rm i}}{2}\sum_{\alpha=1}^M r_\alpha \parallel A_\alpha\psi\parallel ^2 ) |\psi\rangle,
\end{eqnarray}
and $\hat H\equiv H-({{\rm i}}/{2})\sum_{\alpha=1}^M r_\alpha A_\alpha^\dag A_\alpha $ is the non-Hermitian Hamiltonian. In the second integral, $W[\phi|\psi]$ is equal to
\begin{eqnarray}
\sum_{\alpha=1}^M w_\alpha[\phi|\psi_\alpha]\delta\left[\psi_\alpha-\psi\right],
\end{eqnarray}
where $\delta[$ $]$ denotes the Dirac functional and $w_\alpha[\phi|\psi_\alpha]= r_\alpha \parallel A_\alpha|\phi\rangle \parallel ^2$ is the rate of quantum jump of the wave function from $|\phi\rangle$ to the target $|\psi_\alpha\rangle$ due to the action of the jump operator $A_\alpha$. For the precise definition of the target wave function, see Eq.~(\ref{targetwavefunction}) below. A collection of Eqs.~(\ref{densitymatrixzz}) and~(\ref{Liouvillemasterequationquantum}) is called the unraveling of the quantum master equation~(\ref{quantummasterequation}).

The strong similarity of Eqs.~(\ref{Liouvillemasterequationquantum}) and~(\ref{Liouvillemasterclassicalsystem}) arises from the fact that the former also has an interpretation of PDP about stochastic wave functions. As in the case of the classical system, these quantum PDPs are also composed of deterministic pieces and stochastic instantaneous jumps: the deterministic pieces are the solutions of the nonlinear Schr$\ddot{o}$dinger equation,
\begin{eqnarray}
\label{nonlinearSchrodingerequation}
\frac{{\rm d}}{{\rm d}\tau}|\psi(\tau)\rangle &=&-{\rm i}G[\psi(\tau)],
\end{eqnarray}
while the quantum jumps are the instantaneous collapses of the wave functions as
\begin{eqnarray}
\label{targetwavefunction}
|\psi(\tau)\rangle \rightarrow |\psi_\alpha\rangle=\frac{A_\alpha |\psi(\tau)\rangle}{\parallel A_\alpha |\psi(\tau)\rangle \parallel},
\end{eqnarray}
and the rate of the quantum jump is nothing but $w_\alpha[\psi(\tau)|\psi_\alpha]$. Eq.~(\ref{targetwavefunction}) defines the target wave function. {At first sight, $|\psi_\alpha\rangle$ appears $\psi(\tau)$-dependent. However, the jump operators map the latter to special wave functions which are preset by the jump operators themselves. Hence, it is nature to call $\alpha$-jump if a quantum collapse happens because of the jump operator $A_\alpha$. We need point out that in the quantum case it is common that the same target wave function is approached by different types of quantum jumps, e.g., multi-level quantum systems~\cite{Menczel2021}. The textbook by Breuer and Petruccione~\cite{Breuer2002} contains a detailed account of Eqs.~(\ref{densitymatrixzz})~-(\ref{targetwavefunction}).

Based on the definition of the quantum PDPs, we easily see that previous arguments and results about the classical PDPs can be extended to the quantum case. In fact, most of the efforts are simple replacements of notations (an exception is the quantum version of Eq.~(\ref{subLiouvillemasterequation}); see Appendix (A). For instance, the target state $x_\alpha$ is replaced by the target wave function $|\psi_\alpha\rangle$, the notation of the classical trajectory $X_T$ is changed to $\Psi_T$, and the classical flow $\phi_\alpha(\tau)$ is updated to $\hat U(\tau)|\psi_\alpha\rangle$, where $\hat U(\tau)=\exp(-{\rm i}\tau G)$ is the time-evolution operator. Significantly, the KUR~(\ref{generalizedTURPDPs}) and its proof in the quantum regime are exactly the same as those in the classical regime. The cause for this has been explained at the beginning of this section. On the other hand, we also understand why Carollo et al.~\cite{Carollo2019} can obtain the KUR of the Markovian open quantum systems without firstly studying the classical counterparts.

Before closing this discussion, we mention that an important and physically relevant observable is the stochastic heat along quantum trajectories~\cite{Breuer2003,DeRoeck2006,Horowitz2012,Hekking2013, Liu2016a,Liu2018}. For this case, $\omega_\alpha$ in Eq.~(\ref{stochasticcurrent}) represents the energy quantum that is absorbed from or released to heat baths of quantum systems.

\subsection{A resonant two-level quantum system }
Let us illustrate the quantum results by a standard model of quantum optics: in a vacuum, a two-level atom is driven by a resonant field~\cite{Mollow1975,Mandel1995,Breuer1997a}. The quantum master equation of the system {in interaction picture} is
\begin{eqnarray}
\label{QMEtwolevel}
\partial_t\rho(t)=-{\rm i}\left[ H,\rho(t)\right ] + \gamma[\sigma_-\rho(t)\sigma_+ -\frac{1}{2}\{\sigma_+\sigma_-, \rho(t) \}   ],
\end{eqnarray}
where $H=-\Omega(\sigma_- +\sigma_+)/2$ represents the interaction Hamiltonian between the system and the resonant field, $\Omega$ is the Rabi frequency, and $\gamma$ is the spontaneous emission rate. Compared with Eq.~(\ref{quantummasterequation}), we have $M=1$, $r_1=\gamma$, $A_1=\sigma_-$, $A_1^\dag=\sigma_+$, where $\sigma_{+/-}$ are the raising and lowering Pauli operators, respectively. In particular, the target wave function is unique and equal to the ground state: $\psi_1=|g\rangle$. The notion of the PDPs of the two-level quantum system is present if a photon counter continuously records the arriving photons~\cite{Srinivas1981,Carmichael1989,Wiseman1993,Breuer2002,Wiseman2010}. The solution of Eq.~(\ref{nonlinearSchrodingerequation}) is
\begin{eqnarray}
|\psi(\tau)\rangle =\frac{{\rm e}^{-{\rm i}\tau \hat H}|\psi_1\rangle}{\parallel {\rm e}^{-{\rm i}\tau \hat H}|\psi_1\rangle \parallel },
\end{eqnarray}
where the non-Hermitian Hamiltonian is
\begin{eqnarray}
\label{nonHermitianHamiltonian}
\hat H=-\Omega(\sigma_-+\sigma_+)/2 -{\rm i}\gamma\sigma_+\sigma_-/2,
\end{eqnarray}
and the initial wave function is set $|\psi_1\rangle $. Considering the case of $2\Omega>\gamma$, because of the simplicity of the system, we easily obtain~\cite{Breuer1997a}
\begin{eqnarray}
\label{S1}
S_1(\tau)
&=&{\rm e}^{-\int_0^\tau {\rm d}s w[\psi(s)|\psi_1]}=\parallel  {\rm e}^{-{\rm i}\tau \hat H}|\psi_1\rangle \parallel^2
\nonumber \\
&=&{\rm e}^{-{\gamma t}/{2}}\left[1+\frac{\gamma^2}{8\mu^2}\sin^2(\mu t)+\frac{\gamma}{2\mu}\sin(\mu t)\cos(\mu t)\right],\\
\label{p11}
p_{1|1}(\tau)&=&S_1(\tau)w[\psi(\tau)|\psi_1]=r_1\parallel A_1 {\rm e}^{-{\rm i}\tau \hat H}|\psi_1 \rangle \parallel^2 \nonumber \\
&=&\gamma\frac{\Omega^2}{4\mu^2}{\rm e}^{-\gamma \tau/2}\sin^2(\mu \tau) ,
\end{eqnarray}
where $\mu=\sqrt{\Omega^2-r^2/4 }/2$. Eq.~(\ref{S1}) explicitly shows that the distribution of survival time does not decay exponentially as in the pure jump processes~\cite{Gingrich2014,Barato2015a}. This manifests the quantum antibunching effect.

With the above preparations, we study two questions. First, we verify the quantum version of the probability density function in the steady state Eq.~~(\ref{stationarydensity}), which is simply obtained by a replacement of $x$ by $\psi$. Carollo et al. presented this formula without detailed explanation~\cite{Carollo2019}. Because the jump operator is unique, the transition probability $P_{1|1}$ of the Markov chain embedded within these quantum PDPs is trivially equal to $1$. According to Eq.~(\ref{ratealpha}), the rate of jumping to the ground state is
\begin{eqnarray}
\label{c1}
c_1=\frac{1}{\tau_{1}}=\frac{\gamma\Omega^2}{\gamma^2+2\Omega^2},
\end{eqnarray}
where $\tau_1$ is the average survival time in which the system starts with the ground state. Combining Eq.~(\ref{densitymatrixzz}) with the quantum version of Eq.~(\ref{stationarydensity}), and using the property of the Dirac functional, we have
\begin{eqnarray}
\label{stationarydensitymatrixtwolevelatom}
\rho_\infty&=&\sum_{\alpha=1}^{M}c_\alpha\int_0^\infty  {\rm d}\tau {\rm e}^{-{\rm i}\tau \hat H}|\psi_\alpha\rangle \langle \psi_\alpha|e^{{\rm i}\tau \hat H} \\
&=&\frac{I}{2} - \frac{\gamma^2}{\gamma^2+2\Omega^2} \frac{\sigma_z}{2} +  \frac{{\rm i}\Omega \gamma}{\gamma^2+2\Omega^2} \sigma_+  -\frac{{\rm i}\Omega \gamma}{\gamma^2+2\Omega^2} \sigma_-,
\end{eqnarray}
where $I$ and $\sigma_z$ are the identity matrix and the Pauli matrix, respectively. The second equation is a consequence of the substitution of Eqs.~(\ref{nonHermitianHamiltonian}) and (\ref{c1}). We find that the expression of $\rho_\infty$ is consistent with that calculated by the conventional method, which takes the left-hand side of Eq.~(\ref{QMEtwolevel}) as zero and solves a matrix equation. The second problem is to verify the inequality~(\ref{upperboundofrate}). We can analytically calculate the average and variance of the survival time:
\begin{eqnarray}
\tau_{1|1}&=&\tau_1\\
\sigma_{1|1}^2&=&\frac{1}{16\mu^2\Omega^4\gamma^2}(16\Omega^6-12\Omega^4 \gamma^2+6\Omega^2\gamma^4-\gamma^6).
\end{eqnarray}
Because $2\Omega>\gamma$, we can prove that $\tau_{1|1}^2> \sigma_{1|1}^2$, which agrees with the fact that $p_{1|1}(\tau)$ is not an exponential decay function.  Obviously, the antibunching effect leads to a more evenly distributed quantum jumps. In Fig.~(\ref{fig2}) we plot the rate function $I(J)$ for a stochastic counting observable $J[\Psi_T]=k_1[\Psi_T]$ and the local quadratic upper bound under a set of parameters. Appendix~\ref{Appendix1} explains the computational details. We see that the inequality holds. However, we must emphasize that this inequality is usually restricted in a finite range around $J_s$, while these ranges depend on the concrete parameters (data not shown). In contrast, because the KUR~(\ref{generalizedTURPDPs}) is proven under the local condition, it is universal.
\begin{figure}
\includegraphics[width=1.\columnwidth]{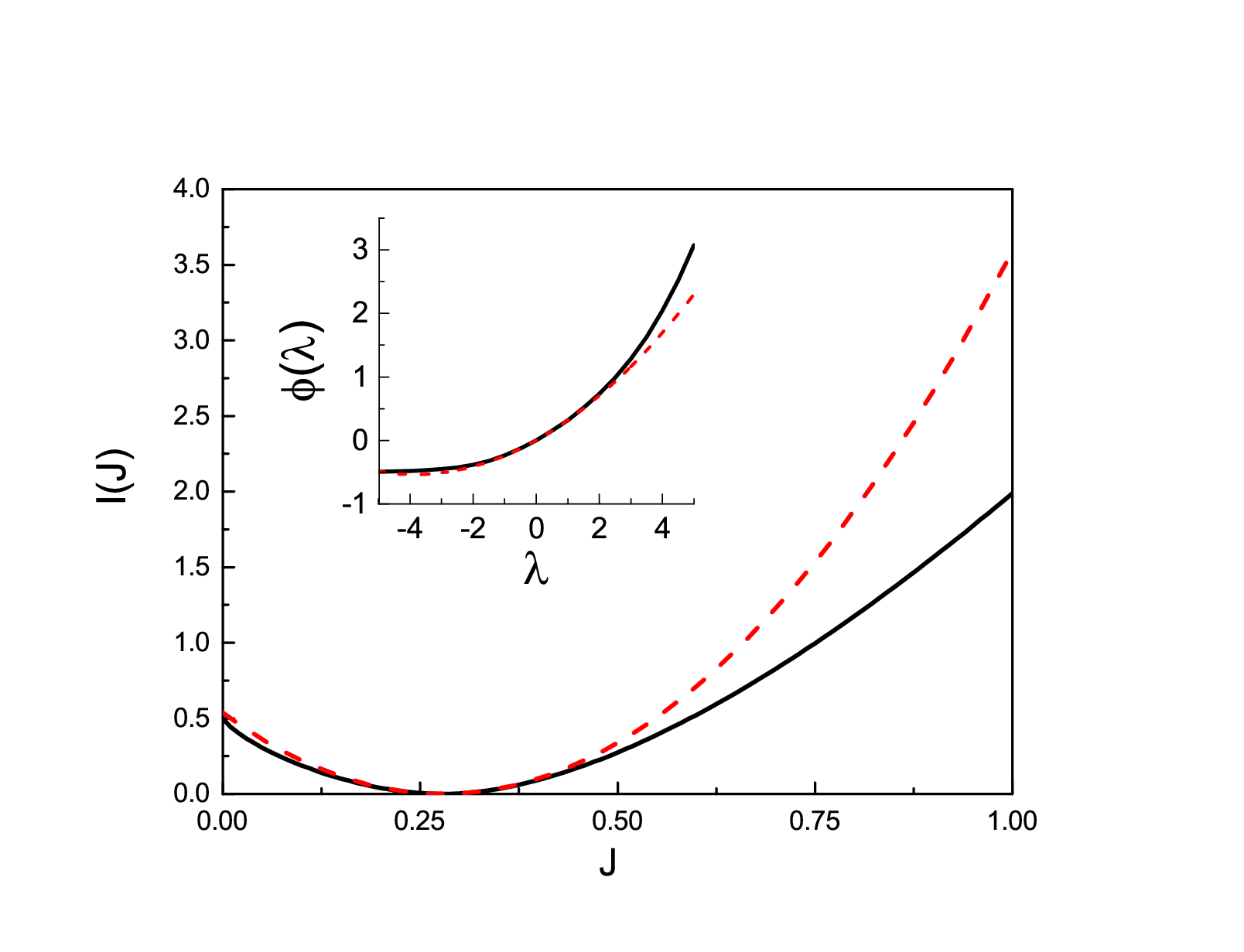}
\caption{The bold and dashed lines are the rate function of the stochastic counting observable and its upper bound, respectively. The inset shows their corresponding scaled cumulant generating functions. Note that the upper bound of the rate function changes to the lower bound of the scaled cumulant generating function. The parameters used here are $\gamma=1$ and $\Omega=0.8$.   }
\label{fig2}
\end{figure}

\section{Conclusion}
\label{section6}
In this paper, we investigated the derivation of the KUR originally found in open quantum systems. We find that this relation is in fact generally valid for the stochastic systems, either classical or quantum, that are described by the PDPs. The underlying cause is that the rates of jumps, rather than the types of the stochastic variables, determine the structure of fluctuations of these stochastic processes.

There are two possible questions worthy of further study. One question is to apply the KUR in practical classical systems that are described by the PDPs and are especially interesting in physics. One potential candidate is a certain type of molecular motors in biological physics: These macromolecules can be described by fully coupled discrete chemical states and continuous mechanical states, or so-called power-stroke model. The second question of further study is relevant to open quantum systems. Hasegawa has obtained another quantum uncertainty relation for the same Markovian open quantum systems. Clarifying their connections should be of interest.

\begin{acknowledgments}
We thank Dr. Carollo for his discussions during this work. This work was supported by the National Natural Science Foundation of China under Grant No. 12075016 and No. 11575016.
\end{acknowledgments}

\appendix

\section{Quantum version of Eq.~(\ref{subLiouvillemasterequation})}
Although the probability meaning of Eq.~(\ref{probrelationbetweentwotimes}) remains invariant in open quantum systems, the mathematical notations therein require replacements: $x\rightarrow\psi$, $y\rightarrow\varphi$, $dy\rightarrow {\rm D}\varphi {\rm D}\varphi*$, and the Dirac function is changed to the Dirac functional. The final result is as follows:
\begin{eqnarray}
P[\psi,t+h;\psi_\alpha,\tau+h]=\int {\rm D}\varphi  {\rm D}\varphi^*  P[\varphi ,t;\psi_\alpha,\tau]\left(1-\Gamma[\varphi]h\right)\delta[\psi-\varphi+{\rm i} G[\varphi]h]+{\it o}(h),
\end{eqnarray}
where we use capital $P$ instead of previous $p$ to denote the meaning of the probability functional.
Analogous to the classical case, we expand both sides of the equation in terms of $h$ until its first order:
\begin{eqnarray}
&&P[\psi,t;\psi_\alpha,\tau]+h\partial_t P[\psi,t;\psi_\alpha,\tau]+h\partial_\tau  P[\psi,t;\psi_\alpha,\tau]\nonumber \\
&=&\int {\rm D}\varphi {\rm D}\varphi^*  P[\varphi ,t;\psi_\alpha,\tau] \delta[\psi-\varphi]-h\int {\rm D}\varphi  {\rm D}\varphi^*  P[\varphi ,t;\psi_\alpha,\tau]\Gamma[\varphi]\delta[\psi-\varphi]+\nonumber \\
&&{\rm i}h \int {\rm D}\varphi  {\rm D}\varphi^*  p[\varphi ,t;\psi_\alpha,\tau]\left\{\int {\rm d}z\left(\frac{\delta}{\delta \psi(z)}\delta[\psi-\varphi]\right) G[\varphi](z)\right.-\nonumber \\
&&\left.\int {\rm d} z \left(\frac{\delta}{\delta \psi^*(z)}\delta[\psi-\varphi]\right) G^*[\varphi](z)\right\}.
\end{eqnarray}
Letting $h$ tend to zero and using the property of the Dirac delta functional, we obtain
\begin{eqnarray}
\label{quantumsubLiouvillemasterequation}
&&\partial_t P[\psi,t;\psi_\alpha,\tau]+\partial_\tau  P[\psi,t;\psi_\alpha,\tau]\nonumber \\
&=&{\rm i}\int {\rm d}z \frac{\delta}{\delta \psi(z)} P[\psi ,t;\psi_\alpha,\tau]   G[\psi](z)-{\rm i} \int {\rm d}z \frac{\delta}{\delta \psi^*(z)}  P[\psi ,t;\psi_\alpha,\tau]   G[\psi]^*(z) -\nonumber \\
&& \Gamma[\psi] P[\psi ,t;\psi_\alpha,\tau].
 \end{eqnarray}
Eq.~(\ref{quantumsubLiouvillemasterequation}) is exactly what we want. Based on this equation, we can derive the quantum Liouville-master~(\ref{Liouvillemasterequationquantum}) in a way similar to that of Eq.~(\ref{Liouvillemasterclassicalsystem}). We do not repeat these details here.

\section{ Calculating rate function of two-level quantum system}
\label{Appendix1}
In Markovian open quantum systems, the most efficient method to calculate rate functions of time-integrated stochastic observables is to solve the tilted (or generalized) quantum master equation~\cite{Esposito2009,Garrahan2010,Liu2016a,Liu2018}. Let the moment generating function (Laplace transform) of the stochastic counting observable be
\begin{eqnarray}
\label{momentgeneratingfunction}
M(\eta,T)=\sum_{N=0}^\infty {\rm e}^{\eta N} P_N(T),
\end{eqnarray}
where $P_N(T)$ is the probability of $N$ counts or jumps of the wave function during time interval $T$. Note that $N=k_1T$. Using the notion of quantum trajectory~\cite{Liu2016a,Liu2018}, we can calculate the function by $M(\eta,T)=Tr[\hat\rho(T)]$, where $\hat\rho(t)$ is the solution of the tilted quantum master equation:
\begin{eqnarray}
\label{tiltedQMEtwolevel}
\partial_t\hat \rho(t)&=&-{\rm i}\left[ H,\hat\rho(t)\right ] + \gamma\left[ {\rm e}^{\eta}\sigma_-\hat\rho(t)\sigma_+ -\frac{1}{2}\{\sigma_+\sigma_-, \hat\rho(t) \}\right]\nonumber \\
&\equiv& {\cal L}_\eta[\hat\rho](t).
\end{eqnarray}
The initial condition is the reduced density matrix of the quantum system at time 0. To calculate the rate function at the long time limit, we use the G$\ddot{a}$rtner-Ellis theorem~\cite{Touchette2008}:
\begin{eqnarray}
I(J)=\max(J\eta-\phi(\eta)),
\end{eqnarray}
where $\phi(\eta)$ is the scaled cumulant generating function and is related to the moment generating function as
\begin{eqnarray}
\label{scaledcumulantgeneratingfunction}
\phi(\eta)=\lim_{T\rightarrow\infty } \ln M(T,\eta)/T.
\end{eqnarray}
To further calculate $\phi(\eta)$, we make use of the well-established result that the scaled cumulant generating function is equal to the maximum eigenvalue $\Lambda_
\eta$ of the tilted quantum master equation~\cite{Lebowitz1999,Esposito2009,Garrahan2010,Carollo2018,Liu2021}, that is,
\begin{eqnarray}
{\cal L}_\eta [\hat\rho_m ]=\Lambda_\eta \hat\rho_m,
\end{eqnarray}
where $\rho_m$ is the corresponding eigenvector. Surprisingly, even for the simple two-level quantum system, it is difficult to obtain a concise analytical expression for $\Lambda_\eta$. We have to use a numerical scheme to solve for the eigenvalues at different $\eta$.

We close this Appendix by clarifying a connection between Eq.~(\ref{tiltedQMEtwolevel}) and a special $P_N(T)$. Assuming that the two-level system starts with the ground state $|g\rangle$, the Laplace transform of $P_N(T)$ over time $T$ has been found to be~\cite{Mollow1975,Mandel1995, Breuer1997a}
\begin{eqnarray}
\label{LaplaceTransfPN}
\hat P_N(\theta)=\frac{1-\hat p(\theta)}{\theta}\hat p(\theta)^N,
\end{eqnarray}
where $\hat p(\theta)$ is the Laplace transform of $p_{1|1}(\tau)$ over $\tau$:
\begin{eqnarray}
\hat p(\theta)=\frac{\gamma\Omega^2/2}{(\theta+\gamma/2)((\theta+\gamma/2)^2+4\mu^2)}.
\end{eqnarray}
Eq.~(\ref{LaplaceTransfPN}) can be obtained by the probability formula of the quantum trajectory~(\ref{probdistributiontraj})~\cite{Breuer1997a}. Let us perform a Laplace transform of $\hat P_N$ over $N$ again:
\begin{eqnarray}
\label{doubleLaplaceTransforms}
\sum_{N=0}^\infty  e^{\eta N}\hat P_N(\theta)=\frac{1-\hat p(\theta)}{\theta}\frac{1}{1-{\rm e}^\eta \hat p(\theta)}.
\end{eqnarray}
Compared with Eq.~(\ref{momentgeneratingfunction}), we immediately find that the right-hand side of Eq.~(\ref{doubleLaplaceTransforms}) is nothing but the Laplace transform of $M(\eta,T)$ over time $T$.
We can directly verify this result by solving the Laplace transform of the tilted quantum master Eq.~(\ref{tiltedQMEtwolevel}). Note that here the initial condition is specified as  $\hat\rho(0)=|g\rangle\langle g|$.

%

\end{document}